\documentclass[preprint,prd,tightenlines,showpacs,nofootinbib,eqsecnum,superscriptaddress]{revtex4}

\usepackage{amsmath}
\usepackage{amsfonts}
\usepackage{amssymb}
\usepackage{mathrsfs}
\usepackage{bm}
\usepackage{graphicx} 
\usepackage{hyperref}

\newcommand{\ud}{\mathrm{d}}
\newcommand{\ui}{\mathrm{i}}
\newcommand{\calW}{\mathscr{W}}
\newcommand{\calO}{\mathcal{O}}
\newcommand{\beq}{\begin{equation}}
\newcommand{\eeq}{\end{equation}}

\def\P{{\cal P}}
\def\bk{{\bf k}}
\def\F{{\cal F}}
\def\p{{\bf p}}
\def\k{{\bf k}}
\def\x{{\bf x}}

\begin{document}

\title{Non-Gaussianity in the Cosmic Microwave Background \\ Induced by Dipolar Dark Matter}

\author{Luc Blanchet}\email{blanchet@iap.fr}
\affiliation{$\mathcal{G}\mathbb{R}\varepsilon{\mathbb{C}}\mathcal{O}$, Institut d'Astrophysique de Paris --- UMR 7095 du CNRS, \\ Universit\'e Pierre \& Marie Curie, 98\textsuperscript{bis} boulevard Arago, 75014 Paris, France}

\author{David Langlois}\email{langlois@iap.fr}
\affiliation{Astroparticle \& Cosmologie (CNRS-Universit\'e Paris 7), \\ 10 rue Alice Domon et L\'eonie Duquet, 75205 Paris Cedex 13, France}

\author{Alexandre Le Tiec}\email{letiec@umd.edu}
\affiliation{Maryland Center for Fundamental Physics \& Joint Space-Science Institute, Department of Physics, University of Maryland, College Park, MD 20742, USA}
\affiliation{$\mathcal{G}\mathbb{R}\varepsilon{\mathbb{C}}\mathcal{O}$, Institut d'Astrophysique de Paris --- UMR 7095 du CNRS, \\ Universit\'e Pierre \& Marie Curie, 98\textsuperscript{bis} boulevard Arago, 75014 Paris, France}

\author{Sylvain Marsat}\email{marsat@iap.fr}
\affiliation{$\mathcal{G}\mathbb{R}\varepsilon{\mathbb{C}}\mathcal{O}$, Institut d'Astrophysique de Paris --- UMR 7095 du CNRS, \\ Universit\'e Pierre \& Marie Curie, 98\textsuperscript{bis} boulevard Arago, 75014 Paris, France}

\date{\today}

\begin{abstract}
In previous work [L.~Blanchet and A.~Le~Tiec, Phys. Rev. D \textbf{80}, 023524 (2009)], motivated by the phenomenology of dark matter at galactic scales, a model of dipolar dark matter (DDM) was introduced. At linear order in cosmological perturbations, the dynamics of the DDM was shown to be identical to that of standard cold dark matter (CDM). In this paper, the DDM model is investigated at second order in cosmological perturbation theory. We find that the internal energy of the DDM fluid modifies the curvature perturbation generated by CDM with a term quadratic in the dipole field. This correction induces a new type of non-Gaussianity in the bispectrum of the curvature perturbation with respect to standard CDM. Leaving unspecified the primordial amplitude of the dipole field, which could in principle be determined by a more  fundamental description of DDM,  we find that, in contrast with usual models of primordial non-Gaussianities, the non-Gaussianity induced by DDM increases with time after the radiation-matter equality on super-Hubble scales. This distinctive feature of the DDM model, as compared with  standard CDM, could thus provide a specific signature in the CMB and large-scale structure probes of non-Gaussianity.
\end{abstract}


\maketitle

\section{Introduction}\label{secI}

The concordance cosmological model $\Lambda$CDM, based on a cosmological constant $\Lambda$ and cold dark matter (CDM), is the state-of-the-art of today's cosmology \cite{OsSt.95}. One of the most important successes of $\Lambda$CDM is its ability to reproduce the observed acoustic power spectrum of temperature fluctuations in the cosmic microwave background (CMB) \cite{HuDo.02,Ko.al.11}. However, despite many successes at cosmological scales, the $\Lambda$CDM model encounters some problems at galactic scales (down to a few $k$pc), notably related to rotation curves of galaxies and the Tully-Fisher relation (see Refs.~\cite{SaMc.02,FaMc.12} for reviews). In particular, it has been shown that the challenging observations for CDM at galactic scales can be summarized in a single empirical law, still unexplained today, and dubbed the modified Newtonian dynamics (MOND) \cite{Mi1.83,Mi2.83,Mi3.83}. 

The MOND proposal fostered many attempts at finding a relativistic modification of gravity, \textit{i.e.} an extension of general relativity (see \textit{e.g.} Refs.~\cite{Be.04,Sa.05,Zl.al.07,Ha.al.08}). Alternatively, based on an analogy between MOND and the physics of dielectric media, the concept of \textit{dipolar} dark matter (DDM) was introduced, at the Newtonian level, as a  reinterpretation of MOND in terms of polarizable dark matter particles without a modification of gravity \cite{Bl1.07}. A relativistic model of DDM in standard general relativity was then proposed \cite{BlLe.08,BlLe.09}. In this model the dark matter particles are endowed with a space-like vector field called the ``dipole'' moment. It was shown that the DDM has the potential to reproduce the phenomenology of MOND at galactic scales, and is strictly equivalent to $\Lambda$CDM at the level of first-order cosmological perturbations around a Friedman-Lema\^itre-Robertson-Walker (FLRW) background. Thus, the DDM behaves like ordinary CDM at early cosmological times. Furthermore the model naturally involves a cosmological constant.

In the present paper we study the DDM model \cite{BlLe.08,BlLe.09} up  to \textit{second} order in cosmological perturbation and show that, at this order, the non-linear dynamics of DDM starts departing from that of ordinary CDM. Consequently, we find that DDM predicts an additional contribution to the curvature perturbation, specifically given by the internal energy of the DDM fluid. We analyse this extra contribution in the evolution of the curvature perturbation at second order in perturbation theory within the Langlois-Vernizzi formalism \cite{LaVe.05,LaVe.06}.

Non-Gaussianity of primordial perturbations in general (multi-field) models of inflation has received a lot of attention in the last few years (see \textit{e.g.} Refs.~\cite{Ba.al.04,Ch.10} for reviews). The best current constraints on non-Gaussianity in the CMB are from the WMAP7 data with the interval $-10 < f_{\rm NL}^{\rm local} < +74$ (95\,\% CL) on the usual parameter $f_{\rm NL}^{\rm local}$ that characterizes the amplitude of the simplest type of non-Gaussianity, namely the local shape. This constraint should be improved by the Planck satellite, and may exclude single-field inflation models if a nonzero value is favored by the data. 

Here we point out that alternative models for dark matter, such as DDM, can provide another possible source of non-Gaussianity; in the case of DDM because the dipole field yields a second-order contribution to the curvature perturbation. We compute the bispectrum of the curvature perturbation, limiting ourselves to super-Hubble scales, and find a specific contribution to non-Gaussianity due to the DDM, assuming that the initial statistical distribution of the dark matter dipole moment is Gaussian at early times. We use two methods, one based on the one-loop bispectrum and the vanishing of the dipole moment in the background (that assumption being consistent with our previous perturbative description of DDM); the other method based on the tree-level bispectrum and assuming that there is a small homogeneous background value for the dipole moment. The two methods are found to agree on the result when interpreted in the sense of cosmic variance.  

The amplitude of the DDM non-Gaussianity signal depends on the \textit{a priori} unknown value of the dipole moment at early times. However, we find that in contrast with usual models of primordial non-Gaussianities, where the curvature perturbation is frozen on super-Hubble scales during the standard cosmological era, the amplitude of the non-Gaussianity induced by DDM increases with time after the radiation-matter equality, because the dark matter dipole field  evolves after this epoch. Although a full study of the evolution of perturbations on sub-Hubble scales would be necessary for detailed predictions, we  expect this new type of non-Gaussianity to lead to a very specific signature, which could be identified  by using different cosmological probes  such as the CMB and large-scale structure surveys.

The plan of this paper is the following. In the next Sec.~\ref{secII}, we present the model of dipolar dark matter. In the subsequent Sec.~\ref{secIII}, we give a perturbative description of the dipolar dark matter up to second order in  cosmological perturbations. In Sec.~\ref{secIV}, we compute the evolution of the curvature perturbation up to second order using a standard perturbation around a FLRW background. Finally, we compute and analyze the non-Gaussianity induced by the DDM contribution to the curvature perturbation in Sec.~\ref{secV}. Throughout this paper we used ``geometrized units'' where $G=c=1$.

\section{Model of Dipolar Dark Matter}\label{secII}

The phenomenological model of dipolar dark matter (and dark energy) developed in Refs.~\cite{BlLe.08,BlLe.09} is based on a matter action in standard general relativity, 
\beq\label{S}
	S_\text{DDM} = \int \ud^4 x \, \sqrt{-g} \, L_\text{DDM}[g_{\mu\nu}, J^\mu, \xi^\mu] \, ,
\eeq
which is to be added to the Einstein-Hilbert action of general relativity (without a cosmological constant), and to the actions of all the other matter fields and fluids (such as baryons, photons, neutrinos, etc.), all described in the standard way. The DDM fluid in a space-time with metric $g_{\mu\nu}$ is described by:
\begin{enumerate}
	\item A conserved mass current $J^\mu = \sigma u^\mu$, such that $\nabla_\mu J^\mu = 0$, where $u^\mu$ is the four-velocity normalized according to $u_\mu u^\mu = - 1$, and $\sigma = {\left( - J_\mu J^\mu \right)}^{1/2}$ is the rest mass density;
	\item A dipole moment vector field $\xi^\mu$, which intervenes in the dynamics only through its projection orthogonal to the four-velocity, $\xi_\perp^\mu \equiv \, \perp^\mu_{\phantom{\mu}\nu} \! \xi^\nu$, where $\perp_{\mu\nu} \, = g_{\mu\nu} + u_\mu u_\nu$. From this vector we construct the polarization $\Pi_\perp^\mu = \sigma \xi_\perp^\mu$, \textit{i.e.} the number density of dipoles, and its norm $\Pi_\perp = \sigma \xi_\perp$, with $\xi_\perp = {\left( \perp_{\mu \nu} \! \xi^\mu \xi^\nu \right)}^{1/2} = {\left( g_{\mu \nu} \, \xi_\perp^\mu \xi_\perp^\nu \right)}^{1/2}$.
\end{enumerate}
The Lagrangian describing the dipolar dark matter fluid reads \cite{BlLe.08,BlLe.09}
\beq\label{L}
	L_\text{DDM} = - \sigma + J_\mu \dot{\xi}^\mu - \calW(\Pi_\perp) \, ,
\eeq
where the dot stands for the covariant derivative with respect to proper time, namely $\dot{\xi}^\mu \equiv u^\nu \nabla_\nu \xi^\mu$. Notice that we have $J_\mu \dot{\xi}^\mu=J_\mu \dot{\xi}^\mu_\perp+\nabla_\rho (J^\rho u_\nu \xi^\nu)$, so that the Lagrangian \eqref{L} admits an equivalent form depending only on the orthogonal projection $\xi_\perp^\mu$; this shows that the only dynamical degrees of freedom of the dipole moment $\xi^\mu$ are those of the projection $\xi_\perp^\mu$, which is a \textit{space-like} vector. Notice also that the first term in Eq.~\eqref{L} is the Lagrangian of a pressureless perfect fluid, \textit{i.e.} that of ordinary CDM.

The potential $\calW$ is a function of the norm $\Pi_\perp = {\left( \perp_{\mu \nu} \! \Pi^\mu \Pi^\nu \right)}^{1/2}$ of the polarization field. Its expansion in powers of $\Pi_\perp$ is determined, in the weak-field limit $\Pi_\perp \ll a_0$ only, by the requirement of recovering the phenomenology of MOND in the non-relativistic regime. Up to third order inclusively, it reads
\beq\label{calW}
	\calW(\Pi_\perp) = \frac{\Lambda}{8 \pi} + 2 \pi \, \Pi_\perp^2 + \frac{16 \pi^2}{3 a_0} \, \Pi_\perp^3 + \mathcal{O}(\Pi_\perp^4) \, ,
\eeq
where $\Lambda$ is the cosmological constant, and $a_0$ is the constant MOND acceleration scale,\footnote{Appearing both in the expansion \eqref{calW}, the  constants $\Lambda$ and $a_0$ should be related numerically  in this model, \textit{i.e.} $\Lambda\sim a_0^2$ (in good agreement with observations).} supposed to be universal and measured to the value $a_0 \simeq 1.2 \times 10^{-10} \, \text{m} \cdot \text{s}^{-2}$ \cite{SaMc.02,FaMc.12}. In the strong-field regime $\Pi_\perp\gg a_0$ (but still non-relativistic), the potential $\calW$ can be adjusted so as to recover the ordinary Newtonian limit \cite{BlLe.08}.

Varying the Lagrangian \eqref{L} with respect to the dipole moment $\xi^\mu$, one obtains some non-geodesic equations of motion for the dipolar fluid:
\beq\label{EOM}
	\dot{u}^\mu = -\hat{\xi}_\perp^\mu \, \calW' \, ,
\eeq
with the notations $\calW'\equiv\ud\calW /\ud\Pi_\perp$ and $\hat{\xi}_\perp^\mu\equiv\xi_\perp^\mu/\xi_\perp$. Varying with respect to the conserved current $J^\mu$ is achieved using a convective variational approach (see the Appendix of Ref.~\cite{BlLe.09}). One obtains, after defining $\Omega^\mu \equiv \dot{\xi}_\perp^\mu + u^\mu \left( 1 + 2 \xi_\perp \calW' \right)$,
\beq\label{Edipole}
	\dot{\Omega}_\mu = \frac{1}{\sigma}\nabla_\mu(\calW - \Pi_\perp\calW')-R_{\mu\rho\nu\tau}\,u^\rho \xi_\perp^\nu u^\tau \, ,
\eeq
where the second term in the right-hand-side (RHS), which involves the Riemann tensor, is analogous to the standard coupling to curvature for the motion of particles with spins in an arbitrary curved background. Finally, the stress-energy tensor of the dipolar fluid can be derived by varying the Lagrangian \eqref{L} with respect to the metric, keeping $J^\mu_*=\sqrt{-g}J^\mu$ as independent of the metric (since $J^\mu_*$ is conserved in the ordinary sense, \textit{i.e.} $\partial_\mu J^\mu_* = 0$). One obtains
\beq\label{T}
	T^{\mu \nu} = \Omega^{(\mu} J^{\nu)} - \nabla_\rho \left( \left[ \Pi_\perp^\rho u^{(\mu} - u^\rho \Pi_\perp^{(\mu} \right] u^{\nu)} \right) - \left( \calW - \Pi_\perp \calW' \right) g^{\mu \nu} \, .
\eeq
This stress-energy tensor is conserved, $\nabla_\nu T^{\mu \nu} = 0$, as a consequence of the equations of motion \eqref{EOM} and evolution \eqref{Edipole}.

Up to an hypothesis of ``weak clusterisation'' of dipolar dark matter during the cosmological evolution \cite{BlLe.08}, it was shown that the model reproduces the phenomenology of MOND in the non-relativistic regime, in the sense that the Bekenstein-Milgrom \cite{BeMi.84} modification of the Poisson equation is recovered. The Euclidean norm of the ordinary gravitational field is then given by the derivative of the potential, $g=\calW'(\Pi_\perp)$, and the MOND interpolating function is related to the potential by $\mu=1-4\pi\Phi(g)/g$, where $\Phi(g)$ is the inverse function of $\calW'(\Pi_\perp)$, \textit{i.e.} is such that $\Pi_\perp=\Phi(g)$.
 
\section{Perturbative description up to second order}\label{secIII}

\subsection{Equivalence between DDM and $\Lambda$CDM at first order}

In the context of cosmological perturbations around a FLRW background, the dipole moment, which is a space-like vector field, has to be perturbative in order not to break the isotropy of the background metric.\footnote{In one approach to computing the non-Gaussianities, in Sec.~\ref{tree} below, we shall relax this hypothesis and assume a small homogeneous background value for the dipole moment.} We thus write $\xi_\perp^\mu = \calO(1)$, which from the equation of motion \eqref{EOM}, together with Eq.~\eqref{calW}, gives $\dot{u}^\mu = \calO(1)$. In the following, we will write apart the term corresponding to the minimum of the potential \eqref{calW} and playing the role of the cosmological constant (dark energy). Accordingly, we split the stress-energy tensor \eqref{T} into $T^{\mu \nu} = - \frac{\Lambda}{8 \pi} \, g^{\mu \nu} + T^{\mu \nu}_\text{DDM}$, and focus on the dipolar dark \textit{matter} contribution $T^{\mu \nu}_\text{DDM}$. It was shown in Ref.~\cite{BlLe.08} that it is possible, up to first order in cosmological perturbations, to recast the stress-energy tensor of this fluid in the form of that of a pressureless perfect fluid, by means of a perturbative redefinition of the four-velocity. Indeed, we define as a convenient variable ($\mathscr{L}_u$ denotes the  Lie derivative along $u^\mu$)
\beq\label{defkappa}
	\kappa^\mu \equiv \mathscr{L}_u \xi_\perp^\mu = \dot{\xi}_\perp^\mu - \xi_\perp^\nu\nabla_\nu u^\mu \, ,
\eeq
which is $\calO(1)$ and satisfies $\kappa^\mu u_\mu=-\xi_\perp^\mu \dot{u}_\mu=\calO(2)$. Note that Eq.~\eqref{defkappa} is intended to be (by definition) valid at any perturbation order. Then, we make use of the following redefinitions of the energy density and four-velocity at first order:
\begin{subequations}\label{rho_Ufirst}
\begin{align}
	\rho &= \sigma - \nabla_\nu \Pi_\perp^\nu + \calO(2) \, , \label{defrhofirst} \\
	U^\mu &= u^\mu + \kappa^\mu + \calO(2) \, . \label{defUfirst}
\end{align}
\end{subequations}
Although the new redefined energy density $\rho$ includes a dipolar contribution, it represents more importantly the conserved energy density associated with the new redefined velocity field $U^\mu$, \textit{i.e.} $\nabla_\mu \mathcal{J}^\mu =\calO(2)$ where $\mathcal{J}^\mu\equiv\rho U^\mu$ is the new current. One can then check that the stress-energy tensor of dipolar dark matter becomes identical to that of a cold dark matter fluid:
\beq\label{equivCDM}
	T^{\mu\nu}_\text{DDM}=\rho \, U^\mu U^\nu + \calO(2)=T^{\mu\nu}_\text{CDM} + \calO(2) \, .
\eeq
Therefore, at the level of first-order perturbations, the dipolar fluid (which also
involves  the cosmological constant) is equivalent to the concordance model $\Lambda$CDM.

This equivalence between DDM and $\Lambda$CDM can also be proved directly at the level of the Lagrangian \eqref{L}. Indeed, at first perturbative order, the redefined conserved current $\mathcal{J}^\mu=\rho U^\mu$ is related 
to the original one, $J^\mu = \sigma u^\mu$, by
\beq\label{Jtilde}
	\mathcal{J}^\mu =  J^\mu + \sigma \kappa^\mu - u^\mu \nabla_\nu \Pi_\perp^\nu + \calO(2) \, .
\eeq
Computing the norm of the current \eqref{Jtilde}, say $r \equiv {\left( - \mathcal{J}_\mu \mathcal{J}^\mu \right)}^{1/2}$, and using the normalization condition $u_\mu u^\mu = -1$, we find
\beq\label{r}
	r = \sigma - J_\mu \dot{\xi}_\perp^\mu - \nabla_\mu \Pi_\perp^\mu + \calO(2) \, .
\eeq
We use this result into Eq.~\eqref{L} and discard the covariant divergence $\nabla_\mu \Pi_\perp^\mu$ since it does not affect the dynamics.\footnote{Notice that the second term in the RHS of Eq.~\eqref{r} can easily be transformed, up to a total divergence, into $\sigma \dot{u}_\mu \xi^\mu$. From the equations of motion $\dot{u}^\mu = \calO(1)$, this term would seem to be of second perturbative order. However, one is not allowed to use the equations of motion back into the Lagrangian, so the second term in \eqref{r} is to be viewed as first order when considered in the DDM Lagrangian \eqref{L}.} Finally, neglecting also the terms quadratic and cubic in the polarization $\Pi_\perp = \sigma \xi_\perp$ in the potential $\calW$, we can recast, at first order in the dipole moment, the Lagrangian \eqref{L}--\eqref{calW} of our DDM model as 
\beq\label{eqCDM}
	L_\text{DDM} = - r - \frac{\Lambda}{8\pi} + \calO(2) =  L_{\Lambda\text{CDM}} + \calO(2) \, ,
\eeq
and we see that it agrees with the Lagrangian describing the dark sector of the cosmological model $\Lambda$CDM --- namely a mass term $r$ without interaction and the cosmological constant (neglecting other species than dark matter and dark energy).

\subsection{Energy density, pressure, and anisotropic stresses at second order}

We now extend the previous analysis to second order in cosmological perturbations, where, as we shall see, deviations from CDM start occurring. The explicit form of the quadratic term in the expansion \eqref{calW} of the potential yields $\Pi_\perp \calW' = 4\pi \, \Pi_\perp^2 + \calO(3)$, such that $\calW-\Pi_\perp \calW' = \frac{\Lambda}{8\pi} - 2\pi \, \Pi_\perp^2 + \calO(3)$. From the requirements that $U_\mu U^\mu = -1 + \calO(3)$ and that the current $\mathcal{J}^\mu=\rho\,U^\mu$ be conserved at this order, $\nabla_\mu\mathcal{J}^\mu=\calO(3)$, we find the following second-order accurate expressions [generalizing Eqs.~\eqref{rho_Ufirst}] for the new variables $\rho$ and $U^\mu$ as functions of the initial variables $\sigma$, $u^\mu$, and $\xi_\perp^\mu$:
\begin{subequations}\label{rho_U}
\begin{align}
	\rho &= \sigma - \nabla_\nu \Pi_\perp^\nu - 4 \pi \, \Pi_\perp^2 - \frac{\sigma}{2} \, \kappa^2 + \calO(3) \, , \label{defrho} \\
	U^\mu &= \left( 1 + \frac{4 \pi}{\sigma} \, \Pi_\perp^2 + \frac{\kappa^2}{2} \right) u^\mu + \left( 1 + \frac{\nabla_\nu \Pi_\perp^\nu}{\sigma} \right) \kappa^\mu + \calO(3) \, . \label{defU}
\end{align}
\end{subequations}
Here $\kappa^2 \equiv \kappa_\nu\kappa^\nu$, where $\kappa^\nu$ is defined by Eq.~\eqref{defkappa}, valid to all orders in perturbation theory. We can then perform a standard $3+1$ decomposition of the stress-energy tensor with respect to the redefined four-velocity $U^\mu$, and obtain
\beq\label{T2}
	T^{\mu \nu}_\text{DDM} = \left( \varepsilon + P \right) U^\mu U^\nu + P \, g^{\mu \nu} + \Sigma^{\mu \nu} + \calO(3) \, ,
\eeq
with the following second-order accurate expressions for the energy density $\varepsilon$, pressure $P$, and anisotropic stress tensor $\Sigma^{\mu \nu}$:
\begin{subequations}
\begin{align}
	\varepsilon &= \sigma - \nabla_\nu \Pi_\perp^\nu - 2 \pi \, \Pi_\perp^2 - \sigma \kappa^2 \, , \\
	P &= \frac{2 \pi}{3} \, \Pi_\perp^2 - \frac{\sigma}{3} \, \kappa^2 \, , \label{P} \\
	\Sigma^{\mu \nu} &= - 4 \pi \, \Pi_\perp^{\langle\mu} \Pi_\perp^{\nu\rangle} - \sigma \kappa^{\langle\mu} \kappa^{\nu\rangle} \, . \label{Sigma}
\end{align}
\end{subequations}
The brackets $\langle \, \rangle$ stand for the symmetric and trace-free (STF) part of a given spatial tensor (orthogonal to $U^\mu$), \textit{e.g.} $\kappa^{\langle\mu} \kappa^{\nu\rangle} = \kappa^\mu \kappa^\nu - \frac{1}{3} \kappa^2 \!\! \perp^{\mu \nu}$; from now on $\perp_{\mu\nu} \, = g_{\mu\nu}+U_\mu U_\nu$ is defined with respect to the four-velocity \eqref{defU}. The anisotropic stress tensor $\Sigma^{\mu\nu}$ is orthogonal to the four-velocity $U^\mu$ and traceless, \textit{i.e.} $U_\nu\Sigma^{\mu\nu} = \calO(3)$ and $\Sigma^\mu_{\phantom{\mu}\mu} = \calO(3)$. Notice that the redefinition \eqref{defU} of the four-velocity has allowed us not only to define the associated conserved mass density \eqref{defrho}, but also to cancel any possible second-order ``heat flow'' $Q^\mu$: there is no term of the type $2Q^{(\mu} U^{\nu)}$, with $Q_\mu U^\mu=0$, in Eq.~\eqref{T2}.

Now, we can introduce the ``specific internal energy'' $W$ of the fluid \textit{via} the definition $\varepsilon \equiv \rho ( 1 + W )$. This dimensionless quantity will play an important role in the following; it explicitly reads
\beq\label{W}
	W = \frac{2 \pi}{\sigma} \Pi_\perp^2 - \frac{\kappa^2}{2} + \calO(3) \, .
\eeq
Hereafter we substitute the four-velocity $U^\mu = u^\mu + \calO(1)$ in place of $u^\mu$ in the definition \eqref{defkappa} of $\kappa^\mu$, since the difference will only appear at third order in perturbation theory. The projected conservation of the stress-energy tensor, $U_\mu\nabla_\nu T_\text{DDM}^{\mu\nu} = 0$, yields
\beq\label{evol}
	\dot{\varepsilon} + \Theta \left( \varepsilon + P + \beta \right) = \calO(3) \, ,
\eeq
where $\Theta \equiv \nabla_\mu U^\mu$ is the expansion of the fluid, and $\beta \equiv \Sigma^{\mu \nu} \sigma_{\mu \nu} / \Theta$ the so-called dissipative pressure, with $\sigma_{\mu \nu} = \nabla_{\langle\mu} U_{\nu\rangle}$ the STF shear tensor. The dot now stands for the covariant proper time derivative along the new fluid worldlines with tangent four-velocity $U^\mu$, such that $\dot\varepsilon\equiv U^\nu \nabla_\nu\varepsilon$.  Making use of $\varepsilon = \rho ( 1 + W )$ and the conservation law $\nabla_\mu ( \rho \, U^\mu ) = \calO(3)$, Eq.~\eqref{evol} can equivalently be written as
\beq\label{PplusBeta}
	P + \beta = - \frac{\rho}{\Theta} \, \dot{W} + \calO(3) \, .
\eeq
Using the explicit expressions \eqref{P}, \eqref{Sigma}, and \eqref{W} for $P$, $\beta$, and $W$, one can easily check that \eqref{PplusBeta} holds in the particular case of our dark matter model.

\section{Curvature perturbation up to second order}\label{secIV}

\subsection{Langlois-Vernizzi formalism}\label{secIVsubI}

We now apply the formalism introduced in Refs.~\cite{LaVe.05,LaVe.06} for the evolution of the curvature perturbation. In this formalism, the usual gauge-invariant curvature perturbation of linear perturbation theory is generalized to a covector (or one-form) $\zeta_\mu$, for which an exact evolution equation is found, valid at any perturbative order and at all scales. First, we define the local number of $e$-folds $\alpha$ by $\Theta = 3\dot{\alpha}$, and a derivative operator projected orthogonally to a given four-velocity field, namely $D_\mu \equiv \, \perp_{\mu}^{\phantom{\mu}\nu} \! \nabla_\nu$. The definition of the curvature one-form is then 
\beq\label{defzeta}
\zeta_{\mu} \equiv D_\mu \alpha - \frac{\dot{\alpha}}{\dot{\varepsilon}} \, D_\mu \varepsilon \, .
\eeq
Notice that one could replace the projected derivatives by covariant derivatives $\nabla_\mu$, or even ordinary gradients $\partial_\mu$ in the above formula. Using the  (approximate) conservation law \eqref{evol}, one obtains the following evolution equation:
\beq\label{LV}
	\mathscr{L}_U \zeta_\mu = \frac{\Theta^2}{3 \dot{\varepsilon}} \biggl( D_\mu (P + \beta) - \frac{\dot{P} + \dot{\beta}}{\dot{\varepsilon}} \, D_\mu \varepsilon \biggr) \, .
\eeq
We refer the reader to Ref.~\cite{LaVe.06} for the original derivation of this relation, but we provide in Appendix \ref{appA} an alternative proof based on the Cartan identity. The RHS of \eqref{LV} would vanish in the case of a perfect, pressureless fluid, for which $\beta = P = 0$, so that $\zeta_\mu$ is conserved along the worldlines of CDM particles. In our case, at second order in cosmological perturbations, the difference between cold dark matter and dipolar dark matter is entirely contained in the (standard plus dissipative) pressure term \eqref{PplusBeta}. Now, notice that the projected derivative picks out the spatial derivative at this order, so for instance $D_\mu \varepsilon=\calO(1)$. Making use of the fact that $W = \calO(2)$ already, we obtain
\beq\label{eqmaitresse}
	\mathscr{L}_U \zeta_\mu = \frac{1}{3}D_{\mu}\dot{W} + \calO(3) \, ,
\eeq
which can then be recast, using again $\dot{u}^\mu=\calO(1)$, into a more familiar conservation law for the variable $\widetilde{\zeta}_\mu \equiv \zeta_\mu - \frac{1}{3} D_\mu W$:
\beq\label{cons_law}
	\mathscr{L}_U \widetilde{\zeta}_\mu = \calO(3)\, .
\eeq
One can also introduce a modified local number of $e$-folds, $\widetilde{\alpha} \equiv \alpha - \frac{1}{3} W$, such that the one-form $\widetilde{\zeta}_\mu$ has a definition similar to that of $\zeta_\mu$, namely 
\beq\label{defzetatilde}
\widetilde{\zeta}_{\mu} \equiv D_\mu \widetilde{\alpha} - \frac{\dot{\widetilde{\alpha}}}{\dot{\varepsilon}} \, D_\mu \varepsilon \, .
\eeq
The conservation law \eqref{cons_law} for the one-form $\widetilde{\zeta}_\mu$ defined with respect to $\widetilde{\alpha}$ can be understood as follows: introducing \eqref{PplusBeta} into \eqref{evol}, one gets $\dot{\varepsilon} + (\Theta - \dot{W}) \, \varepsilon = \calO(3)$, which is the continuity equation of a fiducial pressureless perfect fluid of expansion $\widetilde{\Theta} = \Theta - \dot{W}$. This is precisely the expansion associated with the local number of $e$-folds $\widetilde{\alpha}$ --- the definition of $\Theta$ as the expansion of the velocity does not intervene in the derivation of the relation \eqref{PplusBeta}.

\subsection{Curvature perturbation to second order}\label{secIVsubII}

We now apply the results of the calculations given in \cite{LaVe.05} for second-order cosmological perturbations. We use a FLRW background with line element $\ud s^2=a^2(\eta)(-\ud\eta^2 + \gamma_{ij} \ud x^i \ud x^j)$, where $a(\eta)$ is the scale factor and $\gamma_{ij}$ the spatial metric, and we denote by $'$ the derivative with respect to the conformal time $\eta$. Using the conservation law \eqref{eqmaitresse} at first order, namely
\beq
	\zeta_{i}' = \calO(2) \, ,
\eeq
we obtain, at second order, the following expression for $\zeta_\mu$, introducing the velocity perturbation $v^i$ as $\delta U^i \equiv v^i/a$: 
\begin{subequations}
\begin{align}
	\zeta_{0} & = - v^{i}\partial_{i}\zeta + \calO(3) \,,\\
	\zeta_{i} & = \partial_{i}\zeta + \calO(3) \, . \label{zeta_i}
\end{align}
\end{subequations}
Here $\zeta$ is a scalar that generalizes the curvature perturbation scalar, of which the first and second order expressions in terms of the perturbations of $\alpha$ and $\varepsilon$ can be found in Ref.~\cite{LaVe.05}, and are recalled in Appendix~\ref{appB} below. Specializing the expression \eqref{liesecondorder} for the Lie derivative to second order yields, in combination with \eqref{eqmaitresse},
\beq\label{evol_zeta_i}
	\zeta_{i}' + \partial_i \big( v^j\zeta_j \bigr) = \frac{1}{3} \partial_i W' + \calO(3) \, .
\eeq 
Then, using Eq.~\eqref{zeta_i} we get, up to an irrelevant homogeneous term,
\beq
	\zeta' + v^{i}\partial_{i}\zeta = \frac{W'}{3} + \calO(3) \, .
\eeq 
On super-Hubble scales, the term involving a spatial gradient can be neglected, and we find [up to corrections $\mathcal{O}(3)$]
\beq\label{nonconserv}
	\zeta' = \frac{W'}{3} \, .
\eeq
This shows that, on large scales, the only second-order effect of our modified dark matter model, compared to a standard CDM model, is to correct $\zeta$ by the quantity $W/3$. Denoting by $\zeta_\text{CDM}$ the conserved curvature perturbation generated within the standard CDM scenario, we thus conclude
\beq\label{zeta_W}
	\zeta = \zeta_\text{CDM} + \frac{W}{3} \, .
\eeq
The additional effect, where $W$ is given by Eq.~\eqref{W} with \eqref{defkappa}, is the sum of two terms which scale like the norms squared of the space-like dipole moment and its time derivative; schematically $W \sim \sigma \xi_\perp^2 + \dot{\xi}_\perp^2$.

\subsection{Time evolution of the vector perturbations}\label{secIVsubIII}

Since the internal energy $W$ is quadratic in $\xi_\perp^\mu$ and $\kappa^\mu = \mathscr{L}_U \xi_\perp^\mu$, working at second order in perturbation theory allows us to use the first-order results for the evolution equations, as previously derived in Ref.~\cite{BlLe.08}. Following their notations, we define
\beq
	\xi_\perp^\mu \equiv \left( 0,\lambda^i \right) ,
\eeq
with $\lambda^i=\calO(1)$ the perturbative spatial dipole moment. Easy calculations yield the following expressions:
\begin{subequations}
\begin{align}
	\dot{\xi}_\perp^\mu & = \frac{1}{a} \left( 0, \lambda^i{'} + \mathcal{H}\lambda^i \right) + \calO (2)\,, \\
	\kappa^\mu & =  \frac{1}{a} \left( 0, \lambda^i{'} \right) + \calO (2)\,,\\
	\dot{\kappa}^\mu & = \frac{1}{a^2} \left( 0, \lambda^i{''} \right) + \calO (2)\,,
\end{align}
\end{subequations}
where $\mathcal{H}\equiv a'/a$ is the Hubble parameter in conformal time $\eta$, and we recall that ${'} \equiv \partial / \partial \eta$. Using an overbar to denote background quantities, we get from Eq.~\eqref{W}:
\begin{subequations}
\begin{align}
	W &= 2\pi\bar{\sigma}a^2\lambda^i\lambda_i - \frac{1}{2}\lambda^i{'}\lambda'_i + \calO (3)\,, \label{w} \\
	\dot{W} &= \frac{\mathcal{H}}{a} \left( \lambda^i{'}\lambda'_i - 2\pi\bar{\sigma}a^2\lambda^i\lambda_i \right) + \calO (3)\,, \label{wpoint} 
\end{align}
\end{subequations}
where we denote $\lambda_i \equiv \gamma_{ij}\lambda^j$ and $\lambda'_i \equiv \gamma_{ij}\lambda^j{'}$. In order to derive Eq.~\eqref{wpoint}, we used the first-order accurate evolution equation for the dipole moment $\lambda^i$, which takes the remarkably simple form given in Eq.~(3.29) of Ref.~\cite{BlLe.08}:
\beq
	\lambda^i{''} + \mathcal{H}\lambda^i{'} - 4\pi\bar{\sigma}a^2\lambda^i = \calO(2) \, .
	\label{eqevol}
\eeq

This equation can be solved, at first order, using the FLRW equations for the background dynamics. If we consider the background to be successively dominated by radiation and by (dipolar) cold dark matter, the scalings for the background energy density are $\bar{\rho}_r \propto a^{-4} $ for the radiation domination era and $ \bar{\sigma} \propto a^{-3} $ for the matter domination era, which allows us to find the approximate solutions 
\begin{subequations}
\begin{align}
	\lambda^i(\eta,\x) & = A^i(\x) \, I_0 \bigl( \sqrt{6\Omega\eta} \bigr) + B^i(\x) \, K_0 \bigl( \sqrt{6\Omega\eta} \bigr) & \text{(radiation domination)}\, , \label{solrad} \\
	\lambda^i(\eta,\x) & = C^i(\x) \, \eta^2 + \frac{D^i(\x)}{\eta^3} & \text{(matter domination)}\, .
	\label{solmat}
\end{align}
\end{subequations}
Here $I_0$ and $K_0$ are modified Bessel functions of the first and second kind, respectively, and $A^i$, $B^i$, $C^i$, and $D^i$ are (space-dependent) integration constants. We introduced the inverse of a characteristic timescale,
\beq
	\Omega \equiv a_\text{eq} \left( \frac{8 \pi}{3} \bar{\rho}_\text{eq} \right)^{1/2} ,
\eeq
the subscript ``eq'' indicating the time when the background densities of the radiation and matter fluids are equal, and $\bar{\rho}_\text{eq}$ being the common value of the radiation density $\bar{\rho}_r$ and the matter density $\bar{\sigma}$ at equilibrium.

In fact, it is possible to give a fully analytic solution to the case of the mixture of the two fluids. For this purpose, we rewrite the dynamics in terms of the variable $y \equiv a/a_\text{eq}$. The FLRW equations for the background are then given by 
\begin{subequations}
\begin{align}
	\mathcal{H}^2 & = \Omega^2 \, \frac{y+1}{y^2}\,, \label{calH2} \\
	\mathcal{H}' & = - \frac{\Omega^2}{2} \, \frac{y+2}{y^2}\,,
\end{align}
\end{subequations}
where we used the equation of state $P_r = \rho_r/3$ for the radiation. Finally, we get an evolution equation valid in the two regimes, radiation as well as matter eras [hereafter we drop the neglected remainder $\calO(2)$]:
\beq
	2y(y+1)\frac{\partial^2 \lambda^i}{\partial y^2} + (3y+2)\frac{\partial \lambda^i}{\partial y} - 3\lambda^i = 0 \, .
	\label{mesza}
\eeq
This homogeneous, second-order differential equation is called M\'esz\'aros equation. It appears in standard cosmology in the context of the growth of the matter density contrast on sub-Hubble scales; see \textit{e.g.} Ref.~\cite{PeUzEng}. Its most general solution is known analytically, and reads
\beq\label{sol}
	\lambda^i(y,\x) = A^i(\x) \, \lambda^{+}(y) + B^i(\x) \, \lambda^{-}(y) \, ,
\eeq
where $A^i$ and $B^i$ are again (space-dependent) integration constants, while the growing ($+$) and decaying ($-$) modes read
\begin{subequations}\label{sols}
	\begin{align}
		\lambda^{+} & = y + \frac{2}{3} \, , \\
		\lambda^{-} & = 2\sqrt{1+y} - \left( y+\frac{2}{3} \right) \ln\biggl(\frac{\sqrt{1+y}+1}{\sqrt{1+y}-1}\biggr) \, .
	\end{align}
\end{subequations}
One can check that Eqs.~\eqref{sols} reproduce the asymptotic behaviors of the solutions \eqref{solrad} and \eqref{solmat}. In particular, $\lambda^{-}(y)$ behaves like $\ln(y)$ [or $\ln(\eta)$] when $y \rightarrow 0$, and like $y^{-3/2}$ (or $\eta^{-3}$) when $y \rightarrow + \infty$. We thus have an analytic expression for $\lambda(y)$, which we can inject into the formula \eqref{w} for $W$. This immediately gives, \textit{via} Eq.~\eqref{zeta_W}, the time evolution of the curvature perturbation $\zeta$ on super-Hubble scales.

Discarding the decaying mode $\lambda^{-}$, we can write the evolution of the dipole as
\beq
	\lambda^i(y,\mathbf{x}) = \left(1+\frac32 y\right) \lambda^i_*(\mathbf{x}) \, ,
\eeq
where $\lambda^i_*(\mathbf{x})$ is a time-independent vector field corresponding to the value of the dipole long before the equivalence matter-radiation ($y \ll 1$). Substituting into the expression \eqref{w} for $W$, one obtains the time evolution of $W$ as a function of the ``primordial'' dipole $\lambda^i_*$:
\beq\label{W_F}
	W \equiv \F(a) \, \lambda_*^2 \, ,
\eeq
where $\lambda_*^2 \equiv \gamma_{ij} \lambda_*^i \lambda_*^j$, and the evolution factor is given in terms of the variable $y = a / a_\text{eq}$ by
\beq\label{F}
\F=\frac{9}{16} \,\Omega^2 \left(y+2 +\frac{4}{3y}\right) .
\eeq
This quantity diverges when going backwards in time, and we will use it only sufficiently late so that $W$ remains small enough to be treated as a second order perturbation of $\zeta_\text{CDM}$. After equivalence, $\F$ will increase linearly with the scale factor. At the radiation-matter equality, the typical value of the correction $W$ is given by $\sim (\mathcal{H}_\text{eq} \lambda_*)^2$, which is small if the primordial dipole moment is small compared to the Hubble radius at equivalence. 

\section{Statistical description of the vector perturbations}\label{secV}

\subsection{Formalism and notations}

In this section, we examine how statistical fluctuations of the DM dipole could affect the nonlinear cosmological perturbations and therefore leave a potentially observable  signature. To do so, we will use the analysis of \cite{YoSo.08,Di.al.09,Va.al.09,VaRo.10}, which was developed, in the context of the $\delta N$ formalism \cite{St.85,SaSt.96,LyRo.05,Ly.al.05}, to study models where the curvature perturbation $\zeta$ is allowed to depend on a vector field $A_i$. In the present case, we decompose the total curvature perturbation $\zeta$ into the standard CDM contribution and a contribution due to a vector field $A_i$, directly related to the dark matter dipole, so that one can write, up to second order, 
\beq\label{expansion}
	\zeta= \zeta_{\rm CDM} + N^i \, \delta A_i + \frac12 N^{ij} \, \delta A_i \, \delta A_j \, ,
\eeq
where we use implicit summation over repeated indices. 

Let us now characterize the statistical properties of the vector perturbations, following closely the general approach developed in Refs.~\cite{Di.al.09,Va.al.09}. Since the space-like vector field introduced in the present model is \textit{a priori} purely phenomenological, we will simply assume that it is a collection of uncorrelated Gaussian random fields, without trying to determine its spectrum from a more fundamental description.\footnote{Such a description would consist, for instance, in embedding the dipole in an inflationary model and computing its spectrum from the amplification of its quantum fluctuations during inflation, as was done for fundamental vector fields in Ref.~\cite{Go.al.08}.} It is sometimes convenient to decompose the three components of the vector field, in Fourier space, into the familiar longitudinal ($\ell$), left ($L$), and right ($R$) polarizations, according to the expression
\beq
\label{decomposition_polarisations}
	\delta A_i(\eta,\mathbf{k}) \equiv \int \ud^3\mathbf{x}\,  e^{-\ui \mathbf{k}\cdot \mathbf{x}} \, \delta A_i(\eta,\mathbf{x})  = \sum_\alpha \delta A_\alpha(\eta,\mathbf{k}) \, e_i^\alpha(\hat{\mathbf{k}}) \, ,
\eeq
where the polarization vectors read $e_i^L = (1,\ui,0) / \sqrt{2}$, $e_i^R = (1,-\ui,0) / \sqrt{2}$, and $e_i^\ell = (0,0,1)$, with the $\hat{\mathbf{z}}$ direction  aligned with the direction $\hat{\mathbf{k}}$ of the wavevector $\mathbf{k}$. Following the conventions of \cite{Di.al.09}, we let the change $\mathbf{k} \to - \mathbf{k}$ reverse the $\hat{\mathbf{z}}$ and $\hat{\mathbf{x}}$ directions, but not the $\hat{\mathbf{y}}$ direction, such that $e_i^{\alpha*}(\hat{\mathbf{k}})=-e_i^{\alpha}(-\hat{\mathbf{k}})$. Together with the reality condition $\delta A_i^*(\mathbf{k}) = \delta A_i(-\mathbf{k})$, this implies $\delta A_\alpha^*(\mathbf{k}) = - \delta A_\alpha(-\mathbf{k})$.

The random fields $\delta A_\alpha(\mathbf{k})$ are assumed to be Gaussian and statistically isotropic, with no correlation between different polarization modes or with other fundamental fields (such as the inflaton, the curvaton, etc.). The power spectra $P_\alpha(k)$ for each polarization are defined by the relations
\begin{equation}\label{spectre}
	\langle \delta A_\alpha(\bk) \, \delta A_\beta(\bk')\rangle \equiv - (2\pi)^3 \delta(\bk+\bk') \, P_\alpha(k) \,\delta_{\alpha\beta}\, ,
\end{equation}
where $\delta$ is the usual three-dimensional Dirac delta function, $\delta_{\alpha\beta}$ is the Kronecker symbol and $k = \vert \mathbf{k} \vert$. With these conventions, the power spectra $P_\alpha(k)$ are positive-definite. Using the decomposition \eqref{decomposition_polarisations}, the two-point correlation function of the dipole can be written, in Fourier space, as \cite{Di.al.09}
\begin{align}
	\!\!\!\!\langle \delta A_i(\bk)\, \delta A_j(\bk')\rangle &\equiv (2\pi)^3\delta(\bk+\bk') P_{ij}(\mathbf{k}) \cr
	&= (2\pi)^3\delta(\bk+\bk') \bigl[ T_{ij}^{\rm even}(\bk)P_+(k)+\ui\,T_{ij}^{\rm odd}(\bk)P_-(k)+T_{ij}^{\rm long}(\bk)P_\ell(k) \bigr] \, , \quad
\end{align}
with the even-parity, odd-parity, and longitudinal tensors defined by $T_{ij}^{\rm even} = \delta_{ij} - \hat k_i \hat k_j$, $T_{ij}^{\rm odd} = \varepsilon_{ijk} \hat k_k$, and $T_{ij}^{\rm long} = \hat k_i \hat k_j$, respectively. In the previous expressions, we introduced the  usual totally antisymmetric Levi-Civita symbol $\varepsilon_{ijk}$, such that $\varepsilon_{123} = +1$, as well as the power spectra
\begin{equation}\label{P_pm}
	P_\pm = \frac12 (P_R\pm P_L) \, .
\end{equation}
Note that the Lagrangian \eqref{L} of DDM does not break the parity symmetry. If this property is also shared by the Lagrangian describing the dipole at a more fundamental level, then the left ($L$) and right ($R$) power spectra $P_L$ and $P_R$ should  coincide, implying $P_- = 0$. In the following, we shall keep a general $P_-$.

In the rest of this section, we will apply the above formalism to the DM dipole and compute the power spectrum and bispectrum of the curvature perturbation. Since we have considered the dipole as a perturbation with vanishing background value, we have \textit{a priori} $N^i=0$, which implies that the tree-level contributions to the spectrum and bispectrum simply vanish. One must then consider the one-loop contributions, as we do in the next subsection \ref{loop}. However, in the subsequent subsection \ref{tree}, we will instead assume that the dipole can be decomposed into  a small ``homogeneous'' component, corresponding to its average value in our accessible region, and a spatially-dependent component, in which case the term linear in $\delta A_i$ in Eq.~\eqref{expansion} is nonzero. We will check that the tree-level expressions for the spectrum and bispectrum are compatible with the one-loop calculations of Sec.~\ref{loop}, in the sense of cosmic variance. 

\subsection{One-loop contributions to the spectrum and bispectrum}
\label{loop}

Comparing Eq.~\eqref{expansion} with the expressions \eqref{zeta_W} and \eqref{W_F}, one can directly identify the vector degree of freedom $\delta A_i$ with the primordial DM dipole $\lambda^i_*$, which we have treated as a perturbation, so that 
\beq
 \delta A_i= \lambda_{*i} \, , \qquad N^i=0 \, , \qquad N^{ij} = \frac23 \F(a) \, \delta^{ij}\,,
\eeq
where we assume $\gamma_{ij}=\delta_{ij}$ (flat spatial sections) for simplicity. At the tree-level, the total power spectrum is given by
\beq
P_\zeta(\mathbf{k}) = P_{\zeta_{\rm CDM}}(k) + N^i N^j P_{ij}(\mathbf{k})\,,\quad\text{where} \quad \langle  \zeta(\mathbf{k})   \zeta(\mathbf{k'})\rangle\equiv (2\pi)^3 \delta(\bk+\bk') P_\zeta(\mathbf{k}) \, .
\eeq
Because $N^i = 0$ the contribution of the vector field vanishes. The first non-vanishing effect of the dipole appears in the one-loop contribution
\beq
P^{\rm 1-loop}_\zeta(\mathbf{k})= \frac12 N^{ij} N^{kl} \int \frac{\ud^3\p}{(2\pi)^3} \, P_{ik}(\p) P_{jl}(\k-\p)\,.
\eeq
As an illustration, we will assume in the following that  the power spectrum for the dipole is  scale invariant, \textit{i.e.} $P_{ij}(\k)\propto k^{-3}$. Then, the above integral is divergent at the two singular points $\p=0$ and $\p=\k$. These logarithmic divergences can be regularized by introducing a cut-off $L^{-1}$, where $L$ is a length scale which, in practice, can be chosen as the present Hubble radius: $L=H_0^{-1}$. For instance, the singularity at $\p=0$ gives 
\beq 
\int^k_{L^{-1}} \frac{\ud^3\p}{p^3}\,  \P_{ij}(\p)=\frac{4\pi}{3} \left(2\P_+ +\P_\ell\right)\, \ln (kL) \,\delta_{ij}\,,
\eeq
where we denote $\P_{ij}(\k) \equiv P_{ij}(\k) \, k^3/(2\pi^2) $ and similarly for $\P_{+,\ell}(\k)$. This gives the following estimate for the one-loop contribution to the power spectrum:
\beq\label{P_one_loop}
P^{\rm 1-loop}_\zeta(\mathbf{k}) \simeq \frac{8\pi^2}{27}\F^2  \left(2\P_+ +\P_\ell\right)^2 k^{-3} \ln( kL)\,,
\eeq
where we have substituted $N^{ij} = \frac23 \F \delta^{ij}$. This must remain smaller than $P_{\zeta_{\rm CDM}}$ in order to satisfy observational constraints. 

We now turn to non-Gaussianities. The bispectrum of the curvature perturbation, $B_\zeta(\bk, \bk', \bk'')$, corresponds to the three-point correlation function in Fourier space, 
\begin{equation}
	\langle \zeta(\bk) \zeta(\bk') \zeta(\bk'')\rangle=(2\pi)^3 \delta(\bk+\bk'+\bk'') B_\zeta(\bk, \bk', \bk'')\, .
\end{equation}
Similarly to  the power spectrum, the tree-level bispectrum vanishes when $N^i=0$,  whereas the one-loop contribution is given by
\beq
B^\text{1-loop}_\zeta(\bk, \bk', \bk'') = N^{ij} N^{kl} N^{mn} \int \frac{\ud^3\p}{(2\pi)^3} \, P_{ik}(\p) P_{jm}(\k-\p) P_{ln}(\k'+\p) \, .
\eeq
With a scale-invariant  spectrum $\P_{ij}$, the above integral contains logarithmic divergences for the values $\p=0$, $\p=\k$ and $\p=-\k'$. Introducing as before the cut-off $L^{-1}$, one  finds for the bispectrum the approximate expression (valid at leading order when $L\to+\infty$)
\begin{align}
\label{B_one_loop}
B^\text{1-loop}_\zeta(\bk, \bk', \bk'') &\simeq N^{ij} N^{kl}N^{mn} \left(2\P_+ +\P_\ell\right) \, \ln (kL) \, \delta_{ik} P_{jm}(\k) P_{ln}(\k')
\cr
&=\frac{8}{27}\F^3 \left(2\P_+  +\P_\ell\right) \, \ln (kL) \, P_{ij}(\k) P_{ij}(\k')\,.
\end{align}

\subsection{Tree-level spectrum and bispectrum}
\label{tree}

An alternative derivation of the spectrum and bispectrum consists in assuming that the dipole $\lambda^i_*$, which we have considered as purely perturbative so far, can be decomposed into a nonzero homogeneous component and a spatially-dependent component. Physically, the homogeneous component can be interpreted as the background vector in our accessible box of size $H_0^{-1}$ and corresponds to the average field over our observable region in the particular realization associated with our Universe. It has to be  small in order to be consistent with the analysis of Secs.~\ref{secIII} and \ref{secIV}, and to be compatible with current bounds on anisotropy in the CMB.  Substituting the  decomposition
\beq\label{lambda*}
	\lambda^i_*(\mathbf{x}) = \bar{\lambda}^i_* + \delta \lambda^i_*(\mathbf{x})
\eeq
into the formula \eqref{W_F}, the curvature perturbation \eqref{zeta_W} can be written in the form
\beq
	\zeta = \zeta_{\rm CDM} + \frac13 \F \, \bar{\lambda}_*^2 + \frac23 \F \, \bar{\lambda}_*^i \, \delta \lambda_{*i} + \frac13 \F \, \delta \lambda_*^2 \,.
\eeq
The second term is an irrelevant homogeneous contribution that can be discarded, and the above expression is thus of the form \eqref{expansion} with a nonzero linear term:
\beq
	\delta A_i\equiv \delta\lambda_{*i}\,, \qquad N^i = \frac23 \F \, \bar{\lambda}_*^i\,, \qquad N^{ij} = \frac23 \F \, \delta^{ij}\,.
\eeq
Assuming that $\zeta_{\rm CDM}$ and $\delta A_i$ are uncorrelated, the tree-level power spectrum for the curvature perturbation is given by\footnote{The term involving $P_-$ disappears because $N^i N^j$ is symmetric.}
\begin{align}\label{P_zeta_tree}
	P^{\rm tree}_\zeta(\bk) &= P_{\zeta_{\rm CDM}}(k) + N^i N^j \bigl[ T_{ij}^{\rm even}(\bk) P_+(k)+T_{ij}^{\rm long}(\bk) P_\ell(k) \bigr] \nonumber\\
        &\equiv P_\zeta^{\rm iso}(k) \! \left[ 1 + g \, \bigl( \hat{N}^i \hat k_i \bigr)^2 \right] ,
\end{align}
where the angle-averaged, isotropic spectrum $P_\zeta^{\rm iso}(k)$ and the scale-dependent, anisotropy fraction $g(k)$ read (posing $N^2 \equiv N^i N^i$)
\begin{subequations}
	\begin{align}
		P_\zeta^{\rm iso} &= P_{\zeta_{\rm CDM}} + N^2 P_+ \, , \label{Piso} \\
		g &= N^2 \, \frac{P_\ell -P_+}{P_\zeta^{\rm iso}} \, .
	\end{align}
\end{subequations}

Present observations of the CMB temperature fluctuations require $g$ to be small enough, typically  $g\lesssim 0.1$.\footnote{Although an anisotropy has been detected in the WMAP data, the direction of this asymmetry coincides with the direction orthogonal to the ecliptic, suggesting that this is a systematic effect; see Refs.~\cite{Gr.al.10,HaLe.09}.} In our case, consistency with our assumption that the dipole modifies cosmological perturbations only at second order implies that $g$ is small, such that $P_\zeta \simeq P_\zeta^{\rm iso}$, and that  the isotropic spectrum $P_\zeta^{\rm iso}$ is dominated by the standard $\zeta_{\rm CDM}$ contribution, namely $ P_\zeta^{\rm iso} \simeq P_{\zeta_{\rm CDM}}$. Note that in the particular case where $P_\ell=P_+$, the power spectrum  is automatically isotropic. 

Let us now turn to the bispectrum of the curvature perturbation. Since the  coefficient $N^i$ is now nonzero, the tree-level contribution no longer vanishes, and is given by the expression
\begin{align}\label{Btree}
B_\zeta^\text{tree}(\bk,\bk',\bk'') &= N^i N^j N^{kl} \left[P_{ik}(\k) P_{jl}(\k') + 2~{\rm perm.}\right]
\cr
&= \frac{8}{27}\F^3 \, \bar{\lambda}_*^i \bar{\lambda}_*^j \left[ P_{ik}(\k) P_{jk}(\k') + 2~{\rm perm.}\right] .
\end{align}
One can show that this result agrees with the 1-loop contribution of the previous subsection \ref{loop} by recalling that $N^i = \frac23 \F \bar{\lambda}_*^i$ corresponds to the background vector in our accessible box of size $H_0^{-1}$. It can be interpreted as the average dipole field over our observable region.  One can estimate the typical value of this background field --- or rather of the tensor $\bar{\lambda}_*^i\bar{\lambda}_*^j$ which appears in the bispectrum --- by taking the variance matrix 
\beq
	{\langle\!\langle \bar{\lambda}_*^i(\x) \bar{\lambda}_*^j(\x)\rangle\!\rangle}_L \equiv \int_{L^{-1}}^{k_\text{max}} \frac{\ud^3\p}{(2\pi)^3}\,  P_{ij}(\p)
=\frac13 \left(2\P_+ +\P_\ell\right)\, \ln (k_{\text{max}}L) \, \delta_{ij}\,,
\eeq
where our notation for the double brackets ${\langle\!\langle\cdots\rangle\!\rangle}_L$ means an average on Fourier modes larger than $L^{-1}$ (in contrast to the statistical average denoted before with single brackets $\langle\cdots\rangle$), and where $k_\text{max}^{-1}$ corresponds to the smallest accessible length scale. Substituting $\bar{\lambda}_*^i\bar{\lambda}_*^j$ in the bispectrum \eqref{Btree} by the above expression, one recovers the one-loop bispectrum \eqref{B_one_loop} at the leading order when $L\to +\infty$.\footnote{Similarly one can also recover by the same calculation the spectrum \eqref{P_one_loop}.} This shows that the computation based on  a non-zero background value for the dipole moment is also compatible, in the sense of \textit{cosmic variance}, with a zero averaged value for the background dipole moment.

The shape of the bispectrum \eqref{Btree} naturally involves the preferred cosmological direction $\bar{\lambda}^i_*$. In general it is highly anisotropic and quite complicated. In the particular case where  $P_\ell=P_+$ and $P_-=0$, the bispectrum becomes isotropic with local shape:
\beq\label{bispres}
	B_\zeta(\bk, \bk', \bk'') = \frac{8}{27} \F^3\, \bar{\lambda}_*^2 \left[P_+(k) P_+(k')+P_+(k') P_+(k'')+P_+(k'') P_+(k)\right] .
\eeq
It is then entirely characterized by the isotropic spectrum $P_+(k)$,  grows like $\F^3(a)$, and is proportional to the norm $\bar{\lambda}_*^2$ of the primordial homogeneous dipole.

\subsection{Discussion and conclusions}

In order to get an estimate of the amplitude for non-Gaussianities, it is useful to introduce the usual parameter\footnote{The factor $5/6$ in Eq.~\eqref{def_fNL} comes from the original, historical definition of $f_\text{NL}$ as a parameter measuring non-Gaussianity of the local type in the Bardeen gravitational potential $\Phi = \frac{3}{5} \zeta$, according to the formula $\Phi(\mathbf{x}) = \Phi_\text{G}(\mathbf{x}) + f_\text{NL} \left( \Phi_\text{G}^2(\mathbf{x}) - \langle \Phi_\text{G}^2(\mathbf{x}) \rangle \right)$, where $\Phi_\text{G}(\mathbf{x})$ is a Gaussian random field \cite{KoSp.01}.}
\beq\label{def_fNL}
	f_{\rm NL}(\bk, \bk', \bk'')=\frac56\frac{B_\zeta(\bk, \bk', \bk'')}{P_\zeta(\bk) P_\zeta(\bk')+P_\zeta(\bk') P_\zeta(\bk'')+P_\zeta(\bk'') P_\zeta(\bk)}\,.
\eeq
In general $f_{\rm NL}$ has a complicated (anisotropic) structure in Fourier space. To determine its order of magnitude, let us consider the special case $P_\ell=P_+$, where the bispectrum shape is of the local form \eqref{bispres}, and let us assume that $P_+(k)$ is quasi-scale-invariant, like $P_{\zeta_\text{CDM}}(k)$. Then $P_\zeta(k)$ is also quasi-scale-invariant, and we have
\beq\label{fNL}
	f_{\rm NL} \simeq \frac{20}{81}\F^3 \, \bar{\lambda}_*^2 \, \frac{P_+^2}{P_\zeta^2} = \frac{45}{8192}\left(y+2+\frac{4}{3y}\right)^3 \frac{H_{\rm eq}^6\, a_{\rm eq}^6\, \bar{\lambda}_*^2 \, \P_+^2}{\P_\zeta^2} \, ,
\eeq
where we have substituted the explicit expression (\ref{F}) for $\F$, and used ${\cal H}_{\rm eq} = H_{\rm eq} a_{\rm eq}$. In the present model we cannot predict the amplitude of the primordial dipole moment because this would require a fundamental description of the dipole field during inflation. Nevertheless, we can estimate how present and future bounds on the $f_{\rm NL}$ measured in the CMB temperature anisotropies could constrain the model.

It is convenient to introduce  the dimensionless parameter $x$ giving the relative size of the physical dipole moment\footnote{Recall that, at leading order, the amplitude of the physical dipole moment is given by $\xi_\perp^2=a^2\gamma_{ij} \lambda^i \lambda^j$.} with respect to the Hubble radius, at the time of matter-radiation equality:
\beq
	x \equiv {(\xi_\perp H)}_\text{eq} = (a_\text{eq}\bar{\lambda}_*) H_\text{eq} \, .
\eeq
In this notation we readily find that $\overline{W} = \F \, \bar{\lambda}_*^2 \sim y \, x^2$. Next, using Eqs.~\eqref{spectre}--\eqref{P_pm}, we can relate the order of magnitude of the spectrum $\P_+$ to the perturbation of the dipole moment $\delta \lambda_*$. From the equivalence between the two computations performed in Secs.~\ref{loop} and \ref{tree}, the perturbation $\delta \lambda_*$ should be identified, in order of magnitude, with the background value of the dipole moment $\bar{\lambda}_*$, so that 
\beq
	\P_+ \simeq \alpha\, \bar{\lambda}_*^2 \, ,
\eeq
where $\alpha$ is a number of order one.

We finally find that the expression \eqref{fNL}, evaluated at the epoch of last scattering when $y_\text{ls}=a_\text{ls}/a_\text{eq}\simeq 3$, gives 
\beq\label{fNL_bis}
	f_{\rm NL} \simeq 1.5 \times 10^{17} \, \alpha^2\, x^6 \, , 
\eeq
where we have used the CMB normalization for the power spectrum $\P_\zeta\simeq 2.4\times 10^{-9}$. One can also check that a significant value for $f_{\rm NL}$ is marginally compatible with the assumption that  $\overline{W}$  is  very small ($\overline{W}\ll 10^{-5}$), imposed in order to recover the standard results of linear cosmological perturbations. Indeed, since $\overline{W}\sim x^2$, one finds (with $\alpha\sim 1$)
\beq\label{fNL_ter}
	f_{\rm NL} \sim 150 \left(\frac{\overline{W}}{10^{-5}}\right)^3 .
\eeq
As an illustration, a non-Gaussianity coefficient $f_{\rm NL}\sim 10$ would correspond to $\overline{W} \sim 4 \times 10^{-6}$, \textit{i.e.} to a dipole amplitude of order $x \sim 2 \times 10^{-3}$. Note that the dipole contribution to the power spectrum, 
\beq\label{Xi}
	\Xi \equiv N^2 \, \frac{P_+}{P_\zeta} = \frac{4}{9} \F^2 \, \bar{\lambda}_*^2 \, \frac{\P_+}{\P_\zeta} \, ,
\eeq
is constrained to be very small in order to satisfy the constraint on $\overline{W}$:
\beq
	\Xi \simeq 4\times 10^8\,\alpha\, x^4 \sim 0.04\left(\frac{\overline{W}}{10^{-5}}\right)^2 .
\eeq
The anisotropy fraction $g$ is thus negligible if $P_{\ell}$  and $P_+$ are of the same order of magnitude.

Furthermore, we find that the amplitude of the non-Gaussianity \eqref{fNL} increases with time after the radiation-matter equality, roughly like $\F^3 \sim a^3$, because the dark matter dipole itself grows after that epoch. This behavior is in sharp contrast with usual models of primordial non-Gaussianities, where $\zeta$ remains frozen on super-horizon scales during the standard cosmological era.

That untypical prediction could be used to test the possible dipolar nature of dark matter as follows. Non-Gaussianity in the curvature perturbation can be measured \textit{independently} through the bispectrum of the temperature fluctuations in the CMB, as probed by the Planck satellite, and the bispectrum of the mass density distribution, at low redshifts ($z \lesssim 2$) and large scales, as probed by future galaxy surveys (\textit{e.g.} Euclid, SDSS-III) \cite{Li.al.10,Ve.10}. In order to make specific predictions for the relative amplitudes of the non-Gaussianity that could be measured by these various probes, one should extend the analysis of the present work by taking into account the presence of the dipole in the calculation of the transfer functions. 

To summarize, we investigated the model of DDM at the level of second-order cosmological perturbations. We showed that, on large scales, the only effect of the dipole moment is to correct the curvature perturbation generated within the standard scenario, say $\zeta_\text{CDM}$, by a term proportional to the internal energy $W$ of the dipolar fluid, which is quadratic in the dipole moment [see Eq.~\eqref{zeta_W}]. We then derived, by two different methods in Secs.~\ref{loop} and \ref{tree}, the amplitude of the non-gaussianities generated by this specific contribution. Unfortunately, without a fundamental understanding of the origin of the dipole field introduced in this phenomenological model of dark matter, the amplitude of the dipole moment at early cosmological times cannot be determined, leaving the global scaling of the additional signal of non-Gaussianity \eqref{fNL_bis}--\eqref{fNL_ter} unconstrained. However, the property that the (second-order) contribution of the dipole to the curvature perturbation is time-evolving on super-Hubble scales would lead to specific relations between the levels of non-Gaussianities observed by various probes. 

\acknowledgments
L.B. and S.M. acknowledge partial support from the Agence Nationale de la Recherche via the Grant THALES (ANR-10-BLAN-0507-01-02). D.L. acknowledges partial support from the Agence Nationale de la Recherche via the Grant STR-COSMO (ANR-09-BLAN-0157-01). A.L.T. acknowledges support from NSF through Grant No. PHY-0903631 and from the Maryland Center for Fundamental Physics.

\appendix

\section{Alternative derivation of the relation \eqref{LV}}\label{appA}

We provide here an alternative and more compact derivation of the Langlois-Vernizzi relation \eqref{LV}, based on the use of differential forms and the Cartan identity. In the following, boldface letters  indicate differential forms, arrows indicate four-vectors, and the dot  stands for the contraction of a four-vector with the first index of a differential form, while $\bm{\ud}$  represents the exterior derivative operator. Replacing the projected derivatives $D_{\mu}$ by ordinary derivatives $\partial_{\mu}$ in the definition \eqref{defzeta} of $\zeta_{\mu}$, and using Eq.~\eqref{evol} to rewrite the ratio $\dot{\alpha}/\dot{\varepsilon}$, we may express Eq.~\eqref{defzeta} as 
\beq \label{defzetaform}
	\bm{\zeta} = \bm{\ud} \alpha + \frac{\bm{\ud} \varepsilon}{3(\varepsilon+P+\beta)} \, .
\eeq

We will make use of Cartan's identity for the Lie derivative of a differential form, which reads for a four-vector $\vec{V}$ and a differential form $\bm{\omega}$ of any rank,\footnote{A consequence of this identity is the formula $\bm{\ud} (\mathscr{L}_{\vec{V}} \: \bm{\omega}) = \mathscr{L}_{\vec{V}} \, (\bm{\ud \omega})$.}
\beq \label{cartan}
	\mathscr{L}_{\vec{V}} \: \bm{\omega} = \vec{V} \cdot \bm{\ud} \bm{\omega} + \bm{\ud}\bigl( \vec{V} \cdot\bm{\omega} \bigr) \, .
\eeq
We are going to apply this identity to the four-vector $\vec{U}$ and the one-form $\bm{\zeta}$. We first notice that, by construction (see Eq.~\eqref{defzeta} again), $\bm{\zeta}$ is orthogonal to the four-velocity: $\vec{U}\cdot \bm{\zeta} = 0$. We also note that, in the particular case of a pressureless perfect fluid with $P=\beta=0$, $\bm{\zeta}$ becomes an exact one-form, $\bm{\zeta} = \bm{\ud} (\alpha + \frac{1}{3} \ln \varepsilon)$, and we readily obtain that the RHS vanishes. Coming back to the general case, we have
\beq
	\bm{\ud}\bm{\zeta} = \bm{\ud} \biggl( \bm{\ud} \alpha + \frac{1}{3(\varepsilon+P+\beta)} \, \bm{\ud} \varepsilon \biggr) = -\frac{\Theta^{2}}{3\dot{\varepsilon}^{2}} \, \bm{\ud} (P+\beta) \wedge \bm{\ud} \varepsilon \, ,
\eeq
where we used Eq.~\eqref{evol} again. Now, contracting with $\vec{U}$, we obtain the desired relation:
\beq
	\mathscr{L}_{\vec{U}} \, \bm{\zeta} = \frac{\Theta^{2}}{3\dot{\varepsilon}} \biggl( \bm{\ud}(P+\beta) - \frac{\dot{P}+\dot{\beta}}{\dot{\varepsilon}} \, \bm{\ud} \varepsilon \biggr) \, ,
\eeq
which is identical to Eq.~\eqref{LV}.

\section{Expressions for second-order perturbations}\label{appB}

For completeness, we provide here some explicit expressions for the curvature perturbation in terms of first- and second-order quantities in cosmological perturbations. They can be found in Ref.~\cite{LaVe.05}, together with other definitions and explicit expressions for the metric perturbations. We adopt the convention that any quantity $Q$ is to be decomposed in the following way:
\beq\label{devX}
	Q(\eta,\mathbf{x}) \equiv \bar{Q}(\eta) + Q_{(1)}(\eta,\mathbf{x}) + Q_{(2)}(\eta,\mathbf{x}) + \calO(3) \, ,
\eeq
where $\bar{Q}$ stands for the background quantity and $Q_{(1)}$, $Q_{(2)}$ for the first- and second-order perturbations, respectively. Notice the change in convention with respect to Ref.~\cite{LaVe.05}, where the authors keep a factor $1/2$ in front of the second-order perturbation in Eq.~\eqref{devX}.

For a FLRW background, we have $\ud s^2=a^2(\eta)\,(-\ud\eta^2 + \gamma_{ij} \, \ud x^i \ud x^j)$. We recall that $\alpha$ is the integrated expansion, defined by $\Theta=3 \dot{\alpha}$ (see Ref.~\cite{LaVe.05} for an explicit expression in terms of metric and matter perturbations), and that we define $U^{i}_{(1)} \equiv v^{i}/a$; The curvature one-form $\zeta_{\mu}$ vanishes identically in the background, and we find at first order 
\begin{subequations}
\begin{align}
	\zeta_{0}^{(1)} &= 0 \, ,\\
	\zeta_{i}^{(1)} &= \partial_{i}\zeta_{(1)} \, ,
\end{align}
\end{subequations}
with the following definition for the scalar curvature perturbation $\zeta$:
\beq
	\zeta_{(1)} = \alpha_{(1)}-\frac{\bar{\alpha}'}{\bar{\varepsilon}'} \varepsilon_{(1)} \, ,
\eeq
where the symbol $'$ stands for a derivative with respect to the conformal time $\eta$. We see that at first order $\zeta_{0}$ vanishes and that $\zeta_{i}$ can be written as a pure spatial gradient. This is not the case in general, and at second order we get 
\begin{subequations}
\begin{align}
	\zeta_{0}^{(2)} &= -v^{i}\partial_{i}\zeta_{(1)} \, , \\
	\zeta_{i}^{(2)} &= \partial_{i}\zeta_{(2)} + \frac{\varepsilon_{(1)}}{\bar{\varepsilon}'}\partial_i \zeta'_{(1)} \, ,
\end{align}
\end{subequations}
with 
\beq
	\zeta_{(2)} =  \alpha_{(2)}-\frac{\bar{\alpha}'}{\bar{\varepsilon}'} \varepsilon_{(2)} - \frac{1}{\bar{\varepsilon}'} \alpha_{(1)}' \varepsilon_{(1)} + \frac{\bar{\alpha}'}{\bar{\varepsilon}'^{2}} \varepsilon_{(1)} \varepsilon_{(1)}'  + \frac{1}{2\bar{\varepsilon}'} \left( \frac{\bar{\alpha}'}{\bar{\varepsilon}'} \right)' \varepsilon_{(1)}^{2} \, .
\eeq
On large scales, these expressions agree with other expressions used in the litterature for the conserved curvature perturbation; see Ref.~\cite{LaVe.05} for a discussion. We also reproduce here the expression of the Lie derivative of $\zeta_{i}$ along $U^\mu$ at second order, introducing $A$ as the (first-order) perturbation in the $g_{00}$ component of the metric as $g_{00} = -a^{2}(1+2A)$:
\beq \label{liesecondorder}
	\mathscr{L}_{U} \zeta_{i}^{(2)} = \frac{1}{a}\left[ {\zeta_{i}^{(2)}}' - A {\zeta_{i}^{(1)}}' + v^{j}\partial_{j}\zeta_{i}^{(1)} + \zeta_{j}^{(1)}\partial_{i}v^{j} \right] .
\eeq
Since $\zeta_i^{(1)}$ is a pure gradient, the last two terms in the RHS of Eq.~\eqref{liesecondorder} can be combined to give $\partial_i \bigl( v^j \zeta_j^{(1)} \bigr)$, as used in the evolution equation \eqref{evol_zeta_i}.

\bibliography{}

\end{document}